\documentclass{revtex4}
\usepackage{graphicx}
\usepackage[]{epsfig}

\newcommand{\beq}{\begin{equation}}
\newcommand{\eeq}{\end{equation}}
\newcommand{\beqd}{\begin{displaymath}}
\newcommand{\eeqd}{\end{displaymath}}
\newcommand{\beqa}{\begin{eqnarray}}
\newcommand{\eeqa}{\end{eqnarray}}

\newcommand{\e}{\epsilon}
\newcommand{\sign}{{\rm sign}}

\newcommand{\comment}[1]{}

\newcommand{\bk}{\mathbf{k}}

\newcommand{\ttau}{\tilde{\tau}}
\newcommand{\tlet}{\tilde{t}}

\begin{document}
\title{Towards a Theory of the  Glass Crossover}

\author{Tommaso Rizzo$^{1,2}$}
\affiliation{
$^1$ IPCF-CNR, UOS Rome, Universit\`a "Sapienza", PIazzale A. Moro 2, \\
$^2$ Dip. Fisica, Universit\`a "Sapienza", Piazzale A. Moro 2, I-00185, Rome, Italy \\
I-00185, Rome, Italy}

\begin{abstract}
The standard field-theoretical procedure to study the effect of long wavelength fluctuations on a genuine second-order phase transition is applied to the Mode-Coupling-Theory (MCT) dynamical singularity at $T_c$ in the $\beta$ regime.  Technically this is achieved by a dynamical field-theoretical decoration of MCT that can be studied by a loop expansion. An explicit computation shows that at all orders the leading contributions are the same of a dynamical stochastic glassy equation, {\it i.e.} an extension of the standard MCT equation for the critical correlator with local random fluctuations of the separation parameter. It is suggested that the equation is an essential ingredient in the process that turns the singularity at $T_c$ into a dynamical crossover to activated dynamics.
\end{abstract}

\maketitle

Mode-Coupling-Theory (MCT) provides a rather accurate description of the early stages of the dynamical slowing down in super-cooled glass-forming liquids \cite{Gotze09}.
The theory makes various qualitative and quantitative predictions in agreement with experiments, mainly the two-step nature of the relaxation with time correlators developing plateaus. Various quantities are computed rather accurately within the theory, including notably the non-ergodicity parameter and in addition it provides a detailed set of predictions for the behavior of the critical correlators and the dynamical exponents \cite{Gotze85} that reproduce well the numerical data \cite{Nauroth97,Sciortino01,Weysser10}.
The main problem of the theory is that it predicts at a temperature $T_c$ a dynamical singularity that it is not observed in numerical experiments, on the other hand $T_c$ seems rather to mark a dynamical crossover from a relaxational to an activated dynamical regime \cite{Hansen}.
Due to its quantitative success, many believe that the range of validity of MCT can be extended down to $T_c$ and below including some sort of hopping effect leading to activated dynamics, and various efforts in this direction has been made over the years (see {\it e.g.} \cite{Chong08}). In recent times some authors (see \cite{Mallamace11} and references therein) have also suggested that a crossover temperature identified with $T_c$ is the sole relevant temperature for many glassy materials in contrast to approaches that advocate the presence of an ideal phase transition below the glass transition temperature $T_g$ leading to Vogel-Fulcher-Tammann scaling \cite{Adams65,Stillinger88,Kirkpatrick87d}.  

In this work I discuss the application to MCT of the standard procedure to study long-wavelength corrections to a genuine second-order phase transition. This consists in building a dynamical-field theory with the structure of the original theory which is then studied by means of a loop expansion. The perturbative series can then be resummed and turns out to be equivalent to a dynamical stochastic glassy equation, {\it i.e.} an extension of the standard MCT equation for the critical correlator with local random fluctuations of the separation parameter. This equation is valid in the $\beta$ regime while more work is needed in order to characterize the $\alpha$ regime. Nevertheless it suggest that a complete understanding of how the dynamical singularity at $T_c$ is transformed into a dynamical crossover could be attained by considering the effect of long-wavelength fluctuations with no need to put hopping processes on top of the theory by hand.

The central quantity of MCT is the normalized autocorrelation function of density fluctuations at given wave-vector ${\mathbf k}$
\beq
\Phi(k,t)\equiv \langle \delta\rho^*({\mathbf k},t)\delta\rho({\mathbf k},0)\rangle/S(k)
\eeq
where $S(k)\equiv\langle |\delta\rho(k,0)|^2\rangle$ is the static structure factor.
Within MCT dynamical equations for $\Phi(\bk,t)$ are obtained. The key feature of these equations is that below the critical temperature $T_{c}$ they predict that the long-time limit of the correlator is no longer zero (corresponding to the liquid phase) but becomes positive, $\lim{t \rightarrow \infty} \, \Phi(\bk,t)=f(k) \neq 0$, meaning that the system is in a glassy phase.

For temperatures near the critical temperature one identifies the $\beta$-regime corresponding to time-scales over which the correlator is almost equal to $f(q)$. In the liquid phase ($T>T_c$) this regime is followed by the $\alpha$-regime during which the correlator decays from $f(k)$ to zero. In the $\beta$-regime the time-dependence of the correlator is controlled by the following scaling law \cite{Gotze85}:
\beq
\Phi(k,t)=f(k)+|\tau|^{1/2} f_{\pm}(t/\tau_\beta) \, \xi_c^R(k)
\label{scalfor}
\eeq
where $\tau$ is a linear function of $T_c-T$, {\it i.e.} it is negative in the liquid phase and positive in the glassy phase, correspondingly the scaling functions $f_{+}(x)$ is to be used in the glassy phase while $f_-(x)$ has to be used in the liquid phase.
The function $f_{\pm}(x)$ obeys the scale-invariant equation:
\beq
\pm 1=f_{\pm}^2(x)\left(1-\lambda\right) +\int_0^x (f_{\pm}(x-y)-f_{\pm}(x))\dot{f}_{\pm}(y)dy
\label{SVDYN2}
\eeq
For small values of $x$ both the functions $f_{\pm}(x)$ diverge as $1/x^a$, while for large values of $x$ $f_+(x)$ goes to a constant while $f_-(x)$ diverges as $-x^b$ where the exponents $a$ and $b$ are determined by the so-called parameter exponent $\lambda$ according to:
\beq
\lambda={\Gamma^2(1-a)\over\Gamma(1-2a)}={\Gamma^2(1+b)\over\Gamma(1+2b)}
\label{lambdaMCT}
\eeq
The parameter exponent $\lambda$ controls also the time scale of the $\beta$ regime that diverges with $\tau$ from both sides as $\tau_\beta \propto |\tau|^{-1/(2\,a)}$ with an unknown model-dependent factor.  By using matching argument one can also argue that the time-scale of the $\alpha$ regime increases as $\tau_\alpha \propto |\tau|^{-\gamma}$ with $\gamma=1/(2 a)+1/(2 b)$

The above expressions display a great deal of universality, in particular the universal functions $f_\pm(x)$ depend on the model only through the parameter exponent $\lambda$. Note also that the although the order parameter depends on the momentum $k$, the behavior near $T_c$ is controlled solely by the critical mode $\xi_c^R(k)$, meaning that the actual critical quantity is a single scalar, {\it i.e.} the component of $\Phi(k,t)-f(k)$ along the critical mode.

One is therefore interested in developing a field-theory for this scalar field. This theory must certainly include space variations of the field because spatial fluctuations of the order parameter plays a key role in second-order phase transitions, as we will see it is also crucial to include time variations developing a full dynamical fields theory. 
We note that a global order parameter with no space variations cannot account for nucleation phenomena, and in this sense the nature of the MCT equations is essentially mean-field.
The general ideas and motivations for transforming a mean-field theory into a specific field-theory from which Feynman diagrams are generated are rather old in modern physics, see \cite{Franz12} for a detailed discussion in the context of super-cooled liquids. 

The first candidate field-theory to study critical behavior  at $T_c$ is actually a static field theory. This should not be a surprise, after all if $T_c$ marked a true glass transition it should be possible to characterize the system below $T_c$ with a static theory. Less trivial is the fact that the order parameter of the theory is a replicated version of the correlator $\phi_{ab}(x)$. The theory itself is the  following cubic Replica-Symmetric (RS) field theory with $n=1$ replicas:
\begin{widetext}
\begin{equation}
{\mathcal L}={1 \over 2}\int dx \left(- \tau \sum_{ab}\phi_{ab}+{1 \over 2} \sum_{ab} (\nabla
\phi_{ab})^2+m_2\sum_{abc}
\phi_{ab}\phi_{ac}+m_3\sum_{abcd}\phi_{ab}\phi_{cd} \right) -{1 \over
  6}w_1 \sum_{abc}\phi_{ab}\phi_{bc}\phi_{ca}-{1 \over 6}w_2
\sum_{ab}\phi_{ab}^3
\label{T3}
\end{equation}
\end{widetext}
This theory arises naturally in the context of the so-called one-step-Replica-Symmetry-Breaking (1RSB) Spin-Glass (SG) models. Its relevance for structural glasses was originally suggested by the discovery that the critical behavior of these SG systems is controlled by the very same MCT equations  (\ref{SVDYN2}) and (\ref{lambdaMCT}) \cite{Kirkpatrick87c,Crisanti93}. Although the replica method was introduced originally to tackle the problem of quenched disorder we now understands that the replicated order parameter encodes the mean-field physical phenomenon of the breaking of the liquid state into an exponential number of glassy components.
Indeed the replica method can be applied to structural glasses \cite{Mezard99, Franz12} with predictions that are qualitatively similar but quantitatively different from those of MCT.
However the two approaches are both correct in principle \cite{Rizzo13} and the quantitative differences are due to the different approximation schemes used in the computations. This is strongly hinted by the fact that the very same quantitative predictions of MCT can be obtained within the replica method by means of an appropriate approximation scheme \cite{Szamel10}.  
In \cite{Franz12} the standard technique to decorate a replicated  mean-field theory into the field theory (\ref{T3}) are reviewed in details. Essentially the mean-field results is used as an imput for the bare values of the coupling constants. The procedure is then applied to the mean-field predictions obtained within the Hyper-Netted-Chain approximation but one can also use the quantitative values computed within MCT, see \cite{Rizzo13}, or estimated by any other mean. 

In the context of critical phenomena the bare coupling constants of the actual theory are irrelevant because the universal critical exponents do not depend on the their actual values.  However the present theory is not really critical and therefore in the future it may be important to have the best estimates available in order to attain a complete characterization of the MCT crossover.
In three dimensions standard MCT should be definitively the choice because it often provides very good quantitative predictions for non-universal quantities like the ergodicity breaking parameter, the critical mode, the parameter exponent and the critical temperature. Its approximations appear to be not appropriate in high dimensions but on the other hand in the limit of high dimensions an exact mean-field theory can be developed (see \cite{Kurchan13} for hard-spheres systems). 

The action (\ref{T3}) makes sense only in the glassy phase $\tau>0$ where it can be extremized by the a RS field constant in space $\phi_{ab}(x)=\phi$ given by the solution of the equation of state:
\beq
\tau =\left(w_1-w_2\right)\phi^2 \ .
\eeq
One can then study systematically the loop expansion around the mean-field solution. Quite surprisingly it has been recently discovered \cite{Franz11b} that the loop expansion is equivalent at all orders to a stochastic equation.
For instance the thermal average of the order parameter $\phi$ in the glassy phase is given by:
\beq
\langle \phi(x) \rangle = [\phi_{\tau+\epsilon}(x)]_\epsilon
\label{phimedio}
\eeq
where the square brackets mean average with respect to a Gaussian distributed random field $\epsilon(x)$ with variance
\beq
[\epsilon(x)\epsilon(y)]=-4 (m_2+m_3) \delta (x-y) 
\eeq
and $\phi_{\tau+\epsilon}(x)$ is the solution of the following equation:
\beq
\tau + \epsilon(x) =-\nabla^2 \, \phi +\left(w_1-w_2\right)\phi^2(x)
\eeq
Therefore the inclusion of fluctuations leads to a model with local random fluctuation of the temperature (the random field $\e(x)$).
This result poses various problems. First of all for a given realization of the random field there can be more than one solution. This however is not a major problem as one can think of invoking a maximum condition (motivated dynamically) in order to select the relevant solution. But there is evidently a more serious problem: there can be fluctuations of the temperature that drive portion of the system in the liquid phase meaning that the real solution of the stochastic equation disappears. This implies that the whole static construction is inconsistent when fluctuation are considered. However we are happy with this because it implies that there is simply no glass transition at $T_c$, consistently with all expectations.

In order to understand how the transition at $T_c$ becomes a crossover and to characterize it quantitatively one has to reintroduce dynamics into the problem. 
However the insight gained from the static treatment turns out to be fundamental. Indeed it was recently recognized \cite{Parisi13} that there is a close analogy between the static replica theory and the dynamical theory for the critical correlator in the $\beta$-regime.
This is clearly seen if one adopts a superfield description of the dynamics where one can argue that the dynamical field theory of the super-field correlator has the {\it same} structure of the replicated field theory (\ref{T3}) with the {\it same} coupling constants. 
Most importantly when the equation of state for the critical super-field correlator are translated into the those of the correlator one finds that they have precisely the structure of the MCT critical equation (see \cite{Parisi13} sect. III.D):  
\beq
\tau =\left(w_1-{w_2}\right)\phi^2(t)+w_1\int_0^t (\phi
(t-y)-\phi (t))\dot{\phi}(y) dy
\eeq
from which one identifies \cite{calta1}: 
\beq
\lambda={w_2 \over w_1} \ .
\eeq 
In the following to lighten the notation we will assume without loss of generality that $w_1=1$ because this can be always achieved by a change in the normalization of the critical mode $\xi_c^R(k)$.
We have computed (details elsewhere) perturbative loop corrections of the dynamical field theory with the structure (\ref{T3}) around the dynamical solution for the critical correlator in the $\beta$-regime. The first step is the computation of the  the scaling form equivalent to (\ref{SVDYN2}) for the bare propagator. This is asssociated the four-point susceptibilities that have been studied intensively in recent times \cite{Berthier07}.  Then one has to determine the rules to evaluate all possible diagrams from which a mapping to a stochastic equation can be shown at all orders following Parisi and Sourlas
\cite{Parisi79}. The computation is rather complex has some essential features of the replica case with a crucial difference of purely dynamical origin.
In the end the solution is still of the form (\ref{phimedio}) 
\beq
\langle \phi(x,t) \rangle = [\phi_{\tau+\epsilon}(x,t)]_\epsilon
\label{phimediodyn}
\eeq
with the difference that $\phi_{\tau+\epsilon}(x,t)$ is now the solution of the following glassy dynamical stochastic equation:
\beq
\tau  + \epsilon(x) =-\nabla^2 \, \phi(x,t)+\left(1-\lambda\right)\phi^2(x,t)+\int_0^t (\phi
(x,t-t')-\phi (x,t)){d {\phi} \over dt'}(x,t') dt'
\label{stochglass}
\eeq
note that much as eq. (\ref{SVDYN2}) also the above equation is time scale-invariant and for all $x$ the field $\phi(x,t)$ diverges at small times as $1/t^a$. The actual constant is the same for all $x$ but it is non-universal and it is fixed by the microscopic details of the model. In order to fix it one can adopt the convention  \cite{Gotze09} that  $\lim_{t \rightarrow 0} \phi(x,t) t^a=1$.
Another common feature with (\ref{SVDYN2}) is that it is only valid provided $\phi(x,t)$ is small, {\it i.e.} where the correlator is near the non-ergodicity parameter corresponding to the plateau of the correlator. In particular this holds only on the time scale of the $\beta$ regime and for values of the separation parameter $\tau$ not too large in absolute value.
Note however that these conditions are not perfectly well-defined because there is no genuine dynamical singularity.
The glassy stochastic equation has several interesting features:
\begin{itemize}
\item  According to the equation the dynamics in the $\beta$ regime is the average of a collection of solutions of the MCT equation for the critical correlator with a local randomly fluctuating separation parameter. If we identify a solution with a given physical system the physical picture may look odd in the mean-field case (that will be discussed below) because each solution has a single separation parameter. However in finite dimensions different regions in space are uncorrelated and any solution has qualitatively the same behavior in the thermodynamic limit. This behavior is characterized by strong dynamical heterogeneities: local fluctuations of the separation parameter induce {\it also below $T_c$}  localized liquid regions with higher mobility than the remaining part of the system. In these regions the field $\phi(x,t)$ will decrease indefinitively meaning that at some point the theory must be abandoned and the correlator enters the $\alpha$ regime. This implies that the $\beta$ regime is always followed by the $\alpha$ relaxation and therefore the transition is avoided. On the other hand below $T_c$ the separation parameter increases and these liquid regions become increasingly rare marking a crossover to an activated regime.  

\item The fact that the integral over all values of the random temperature in (\ref{phimediodyn}) is well defined depends crucially on the presence of the last term in eq. (\ref{stochglass}). A purely relaxational term of the form $d\phi/dt$ would be a disaster because $\phi(t)$ of a liquid solution would go to minus infinity in {\it finite} time.  

\item If we study equation (\ref{stochglass}) perturbatively in the strength of the temperature fluctuations the solution at leading order is homogeneous in space and will be given by $f_+(t)$ in the (pseudo)-glassy phase $\tau>0$. Therefore in the perturbative loop expansion one will never be able to see that there rare regions of the systems that are above $T_c$ and decay through $f_-$. However when we resum the loop expansion and obtain eq. (\ref{stochglass}) this problem disappears. Therefore the underlying mechanism is both activated (exponentially small probabilities) and non-perturbative (we go from $f_+$ to $f_-$).  
\end{itemize}

A complete description of the dynamics near $T_c$ requires a characterization of the $\alpha$ regime where eq. (\ref{stochglass}) is no longer valid. 
The matching between the $\beta$ and $\alpha$ regimes is not at all trivial. For instance, according to the above equation the local random fluctuations have no time dependence on the time scale of the $\beta$-regime but this cannot be true on the scale of $\alpha$ regime because the regions of greater mobility cannot be the same at all times. On the other hand the equation of the $\beta$ relaxation carries already substantial quantitative information. In order to illustrate this we consider the mean-field case in which we remove the space dependence of the field $\phi(x)$ and of the random temperature $\epsilon(x)$. The resulting theory is of direct relevance for the class of mean-field discontinuous spin-glass models defined on random-lattices \cite{Franz11b}. The equation reduces to the standard equations of the critical correlator of MCT and its solutions can be written in terms of the functions $f_{\pm}$.
The variance of the random field is $O(1/N)$ where $N$ is the number of spins in the system, therefore in the thermodynamic limit we see that at $\tau=0$ we have a true dynamical singularity. However a careful analisys shows  that at any finite $N$ there is 
a critical region of temperatures $\tau=O(N^{-1/2})$ where one sees that the transition is actually avoided. One finds that in the critical  region $\tau=O(N^{-1/2})$ the scale of the $\beta$-regime is $\tau_\beta \equiv N^{1/4a}$ and the scale of correlator is $N^{-1/4}$. 
These dynamical scaling-laws have been already verified numerically \cite{Sarlat09,Franz11b} since they can be derived through matching arguments from the (ill-defined) static treatment of the glassy phase. The full-fledged dynamical treatment is important not only because it allows to obtain the otherwise inaccessible scaling functions but also because the success of such matching arguments depends crucially on the nature of eq. (\ref{stochglass}). Indeed, as we said already, a simple relaxational dynamics would destroy the matching and leads to a completely different behaviour. 

In the following we will concentrate on the critical correlator from which in turn we will extract information on the $\alpha$-regime and the crossover from relaxational to activated dynamics.
In the critical region the solution of the glassy stochastic equation leads to the following expression for the critical correlator:
\beq
\langle \phi(t) \rangle_{\tau} = {1 \over N^{1/4}} \phi^{scal}_{\tau N^{1/2}}(t \, N^{-1/4a})
\label{op1}
\eeq
where  $\phi^{scal}_{\tilde{\tau}}(\tilde{t})$ is a scaling function independent of $N$ defined as:
\beq
\phi^{scal}_{\tilde{\tau}}(\tilde{t}) \equiv [\phi_{\tilde{\tau}+\tilde{\e}}(\tilde{t})]_{\tilde{\e}} \ ,
\eeq
the square bracket mean average with respect to the random Gaussian variable $\tilde{\epsilon}$ that is the rescaled random field $\tilde{\epsilon}=N^{1/2}\epsilon$ and has a finite variance in the thermodynamic limit \footnote{we follow the convention that tilded variables are rescaled variables that remain finite in the thermodynamic limit.}.
The function inside the square brackets  $\phi_{\tilde{\tau}+\tilde{\e}}(\tilde{t})$ is then defined in terms of the critical functions $f_{\pm}$ as:
\beq
\phi_{\ttau}(\tlet) \equiv |\ttau|^{1/2}  \, f_{\sign(\ttau)}(\tlet |\ttau|^{1
    \over 2 a})
\label{phiTAU}
\eeq
As we said before eq. (\ref{stochglass}) is no longer valid in the $\alpha$-regime. Nevertheless assuming a matching between the late $\beta$ and the early $\alpha$ regime we can extract information on crossover from power law to activated dynamics.
Qualitatitely the decay is controlled by the solutions that are in the liquid phase {\it i.e.} those for which $\ttau+\tilde{\e}<0$. These solutions leads to $\phi^{scal}_{\ttau}(\tlet) \propto \tlet^b$ at large values of $\tlet$.  Together with the condition that expression (\ref{op1}) must become $O(1)$ in the late-$\beta$/early-$\alpha$ regime this leads to $1 \propto N^{-1/4}(\tau_\alpha/\tau_\beta)^b$ and therefore
\beq
\tau_\alpha \propto N^{\gamma/2}\ .
\label{alphascal}
\eeq
This expression has been already proposed in \cite{Sarlat09} to explain numerical observations. 
Note that the liquid solutions that drive the decay become less and less probable as we lower the temperature (going to large positive $\ttau$), indeed they must corresponds to a fluctuation of the field of order $\ttau$ that has an exponentially low probability. In mathematical terms this can be quantified studying the dependence on $\ttau$ of the constant in front of (\ref{alphascal}). In order to do this we consider (\ref{phiTAU}) for negative values of $\ttau$ and large values of $\tlet$, this gives:
\beq
\phi_{\ttau}(\tlet) \propto -|\ttau|^{{1\over 2}+{b \over 2 a}}  \tlet^b
\label{CTAULIMIT}
\eeq
times an irrelevant constant independent of $\ttau$.
Therefore we see that average over the solutions can be rewritten (putting for simplicity the variance of the random field to unity) as:
\beq
e^{-\ttau^2/2}\int_0^{\infty}dz e^{- z \ttau-z^2/2}z^{{1\over 2}+{b \over 2 a}}dz \propto e^{-\ttau^2/2} \ttau^{-{3\over 2}-{b \over 2 a}}
\eeq
rescaling as $z \rightarrow z/\ttau$ the matching condition reads:
\beq
1 \propto N^{-1/4} e^{-\tau^2/2} \tau^{-{3\over 2}-{b \over 2 a}} \, (\tau_\alpha/\tau_\beta)^b
\eeq
from which we can exhibit an exponential increase of the relaxation time as we go to large positive values of $\ttau$ (low temperatures) that has to be contrasted to the behaviour in the high temperature region (large negative $\ttau$) where one recovers the MCT result:
\beq 
\tau_\alpha \propto N^{\gamma/2} \, \times  \left\{ \begin{array}{ll}
         e^{\ttau^2 \over 2 b} \ttau^{1/b+\gamma} & \mbox{if $\ttau \rightarrow  \infty$};\\
        |\ttau|^{-\gamma} & \mbox{if $\ttau \rightarrow  -\infty$}.\end{array} \right. 
\eeq 
The behaviour from the high temperature region would predict a dynamical singularity at $\ttau=0$ which is instead turned into a crossover between the standard power-law increase of $\tau_\alpha$ for $T>T_c$ to an activated regime for $T<T_c$.
Note the dependence of the exponent on the inverse of the dynamical exponent $b$.

In the general case the complete characterization of the $\beta$ regime near the dynamical crossover at $T_c$ and the comparison with experimental and numerical data are delicate problems that are left for future work. In this respect we note that the stochastic glassy equation (\ref{stochglass}) is considerably similar to MCT's eq. (\ref{SVDYN2}) and it can be argued that in order to put numbers into it one does not need to go back and forth from MCT to a dynamical field theory and back to eq. (\ref{stochglass}). The values of $\lambda$ and $\tau$ are indeed provided by standard MCT and the variance of the random temperature can be obtained by reading the coefficients $m_2$ and $m_3$ from the expansion of \cite{Rizzo13}. A little bit less trivial is the computation of the coefficient of the Laplacian that should be estimated by means of the inhomogeneous MCT extension discussed in \cite{Biroli06}.  

In the following we will make some mainly qualitative comments on what should happen when space is put back into the problem considering the full equations (\ref{stochglass}). 
In the region of high temperatures (large negative values of $\tau$) we expect $\tau$ to dominate both on the random field and on the gradient term that can be both treated perturbatively. This is the region where the standard MCT scalings apply, in particular $\tau_\alpha \propto |\tau|^{-\gamma}$ and $\tau_\beta \propto |\tau|^{-1/(2a)}$. The dynamical correlation length will also increase with the mean-field exponent $\xi \propto |\tau|^{-1/4}$ in agreement with previous results \cite{Biroli06}.
This state of things changes when the corrections start to be relevant. In dimension $D<8$ this should happen in correspondence to the violation of the Ginzburg criterion (which is the same of the static theory \cite{Franz12}):
\beq
1 \gg \tau^{D-8 \over 4}(m_2+m_3)(1-\lambda)^2  \ .
\eeq
Approaching $\tau=0$ the transition is avoided and $\tau_\alpha$, $\tau_\beta$ and $\xi$ remain finite. Lowering the temperature well below $T_c$, $\tau$ is again large an positive and we enter the activated regime. In this regime the typical solution appears to be frozen; its eventual relaxation is due to rare regions where $\tau+\e(x)$ is negative (corresponding to the liquid) because of fluctuations with exponentially small probability. This implies that the dynamics is extremely heterogeneous in space, a property which is indeed considered  a key feature of glassy dynamics \cite{Kob97,Ediger00}.
On the other hand scaling suggests that the actual size of these regions {\it decreases} in the deep activated regime ($\tau \gg 1$) and one may ask if there is a connection to  observations of a non-monotonous behavior across $T_c$ of a properly defined dynamical correlation length \cite{Kob12}. Clearly the size of the liquid regions cannot decrease beyond the microscopic scale and below a certain temperature the continuous stochastic glassy equation must be abandoned. What happens then requires a different analysis, standard Arrhenius behavior being a possibility.

An important point that one has to realize is that precisely because the transition is avoided there can be no universal crossover function in the standard sense of critical phenomena. The theory described by eq. (\ref{stochglass}) arises as a resummation at all order of the most divergent corrections and as a consequence it is only valid at large distances. 
In the case of an actual second-order phase transition this is consistent because the correlation length grows indefinitely near the transition leading to a decoupling from the microscopic details of the model. In the present theory however the correlation length does not diverge at $T_c$  and therefore the behavior of the system near the (pseudo)-critical temperature will retain, at least in principle, a non-universal dependence on the small-lengthscale details of the system which are not present in eqs. (\ref{stochglass}).  Other interesting questions concern the role of the pseudo-upper critical dimension that in this case is $D=8$ \cite{Franz11b} and the actual size of the region where the present theory holds (maybe down to $T_g$?).
More generally one would like to know if the properties that the theory would have if it was  genuinely critical are completely  wiped out or they leave some trace in the crossover. The answers to these questions are likely to be model-dependent and could possibly be obtained from numerical solution of the stochastic glass equations eventually supplemented  with information on small-scale details of the specific system under study.   

The present theory of the $\beta$ relaxation near the MCT crossover appears in the end conceptually very simple. 
This simplicity however should not deceive. One should remember that it is not a phenomenological extension of MCT designed explicitly to produce some sort of hopping dynamic. The starting point is a field-theoretical decoration of MCT which is the standard procedure to study long wavelength corrections to a genuine second-order transition. 
It is an explicit computation that shows that the result is equivalent {\it at all orders} in perturbation theory to the dynamical glassy stochastic equations for non-trivial technical reasons that have maybe a deeper explanation.

\end{document}